\documentclass[singlecolumn,superscriptaddress,11pt]{revtex4}
\usepackage{amsmath}
\usepackage{latexsym}
\usepackage{amsthm}

\usepackage{amssymb}
\usepackage{graphicx,epstopdf,subfigure,color,hyperref,enumerate}
\usepackage[normalem]{ulem}

\usepackage{xcolor, dsfont, bbm}
\usepackage{pgf}
\usepackage{times}
\usepackage{graphics}
\usepackage{wrapfig}
\usepackage{etoolbox}
\definecolor{Red}{rgb}{1,0,0}

{\catcode`\|=\active
  \gdef\Braket#1{\begingroup
\mathcode`\|32768\let|\BraVert\left<{#1}\right>\endgroup}}
\def\BraVert{\egroup\,\mid\,\bgroup}

\begin{document}
\title{\Large Extraordinary Interactions between Light and Matter
  Determined \\ by Anomalous Weak Values}

\author{Yakir Aharonov}
\email{yakir@post.tau.ac.il}
\affiliation{School of Physics and Astronomy, Tel Aviv University,
  Tel-Aviv 6997801, Israel}
\affiliation{Institute for Quantum Studies, Chapman University,
  Orange, CA 92866, USA}
\affiliation{Schmid College of Science, Chapman University, Orange, CA
  92866, USA}
\affiliation{Iyar, The Israeli Institute for Advanced Research, POB
  651, Zichron Ya'akov 3095303, Israel}

\author{Eliahu Cohen}
\email{eli17c@gmail.com}
\affiliation{Physics Department, Centre for Research in Photonics, University of Ottawa,                                                  Advanced Research Complex, 25 Templeton, Ottawa ON Canada, K1N 6N5}
\affiliation{H.H. Wills Physics Laboratory, University of Bristol, Tyndall Avenue, Bristol
BS8 1TL, UK}
\affiliation{Iyar, The Israeli Institute for Advanced Research, POB
  651, Zichron Ya'akov 3095303, Israel}

\author{Avishy Carmi}
\email{avcarmi@bgu.ac.il}
\affiliation{Center for Quantum Information Science and Technology and
Faculty of Engineering Sciences
Ben-Gurion University of the Negev, Beersheba 8410501, Israel}
\affiliation{Iyar, The Israeli Institute for Advanced Research, POB
  651, Zichron Ya'akov 3095303, Israel}

\author{Avshalom C. Elitzur}
\email{avshalom@iyar.org.il}
\affiliation{Institute for Quantum Studies, Chapman University,
  Orange, CA 92866, USA}
\affiliation{Iyar, The Israeli Institute for Advanced Research, POB
  651, Zichron Ya'akov 3095303, Israel}

\begin{abstract}

Some predictions regarding pre- and post-selected states are
far-reaching, thereby requiring validation with standard quantum
measurements in addition to the customary weak measurements used so
far, as well as other advanced techniques. We go further pursuing this goal, proposing two thought experiments which incorporate novel yet feasible validation methods of unconventional light-matter interactions.
%The results give the impression that the atom was present in the two MZI arms with
%unusual number: -3 on one side and 4 on the other.
An excited atom traverses a Mach-Zehnder interferometer (MZI) under a
special combination of pre- and post-selection.
%The atom is therefore
%retrodicted to have possessed anomalous weak values during its passage
%through the MZI.
In the first experiment, photons emitted by the superposed atom, after being hit by two laser beams, are
individually counted.
%, and the experiment is repeated many times for statistical significance.
Despite the interaction having definitely taken place, as revealed by the atom becoming ground, the
numbers of photons emitted from each arm of the MZI are predicted, at the ensemble level, to be
different from those expected with standard stimulated emission. In the second experiment, the atom
spontaneously emits a photon while still in the MZI. This photon later
serves as a strong measurement of the atom's energy upon hitting a
photographic plate. The experiment is repeated to enable an
interference effect of the emitted photons. Interestingly, the latter gives the appearance
that the photons have been emitted by the atom from a position much
farther from the two MZI arms $L$ and $R$, as if in a ``phantom arm''
$R'$. Nevertheless, their time of arrival is similar to that of
photons coming from $L$ and $R$.
These experiments also emphasize the key role of anomalous weak values in determining light-matter interactions. In fact, they present a straightforward realization of an entity we term counter-particles, namely pre- and post-selected states acting as if they have negative physical variables such as mass and energy. The novel verification methods we suggest for testing these predictions resemble weak measurements in some aspects, yet result from definite atomic transitions verified by the detected photons.

\end{abstract}

\maketitle

The Two-State-Vector Formalism (TSVF) \cite{TSVF1,TSVF2} offers a
simple yet very efficient and fruitful method of studying quantum
phenomena. Classical physics enables prediction of a future state
based on the system's initial conditions. Conversely, one can
retrodict past states on the basis of final conditions. The two
methods are equivalent, hence each is redundant to the other. Not so in quantum mechanics: Prediction alone, using the
pre-selected wavefunction $| \Psi \rangle$, and retrodiction alone, with
the post-selected wavefunction $| \Phi \rangle$, give only partial (and sometimes conflicting)
information due to quantum uncertainty. However, their {\it combination}
in the form of the two-state $\langle \Phi | \; | \Psi \rangle$ gives
much more information \cite{TSVF1,TSVF2}.
This information is available through inference of the weak value
\cite{AAV} of any operator $\hat{A}$,
\begin{equation}
\label{eq:wv}
\langle \hat{A} \rangle_w = \frac{\langle \Phi | A | \Psi \rangle}{\langle
  \Phi | \Psi \rangle}.
\end{equation}
Moreover, when the two boundary conditions markedly differ from one
another, their combined information gives rise to ``anomalous weak values,'' (also called ``superweak values'' \cite{BerrySW1}) i.e., too large/small or even complex
\cite{AR,Jozsa,Yutaka,BerrySW2,Sok}.

These intriguing values have been demonstrated, so far, mainly with the aid of
weak measurements \cite{AAV} (see however a recent claim of strongly measuring weak values \cite{Denk} and the corresponding comment \cite{LevC}). Although being somewhat controversial as an experimental tool \cite{C1,C2,C3}, weak measurements have led to many practical (e.g. \cite{P1,P2,P3,P4,P5}) and conceptual (e.g. \cite{T1,T2,T3,T4,T5,T6,T7,T8,T9})
achievements. Close in spirit to this paper was the demonstration of spontaneous-emission-based weak measurements \cite{Barak}. Yet, since all the above measurements involve weak coupling between
the measuring pointer and measured system, thereby afflicted with
inherent quantum noise, they have sometimes been explained away,
although erroneously (see for instance \cite{R1,R2,R3}), as artifacts of noise
\cite{FC}. The introduction of projective
(``strong'') quantum measurements for the validation of TSVF predictions
\cite{Pigeon,Dis,OT,Shutter} was therefore a major advance, immune to above objections and offering a deeper understanding of the quantum realm.

Following are two experiments of this kind, where curious
predictions of TSVF are verified with the aid of both weak and strong
measurements. These are performed after the measured atom has
definitely changed its state (as indicated by the post-selection). Then the photons that have interacted with the atom are detected, revealing what has occurred between the pre- and post-selection. The
overall effect, like many fundamental quantum mechanical ones, is
proved on the basis of a sufficiently large ensemble. To the best of our knowledge, not only the
effects, but also the validation techniques are novel,
yet highly feasible. We shall nevertheless focus on concepts, rather than technical details.

This paper is organized as follows. We first present the
basic setup and then analyze two different thought experiments, one involving stimulated emission of radiation and the other incorporating spontaneous emission. We then conclude with
some general consequences.

\subsection{Pre- and Post--Selected States of Atoms}

An excited atom traverses an atomic Mach-Zehnder interferometer (MZI)
(Fig. 1) such that its initial spatial superposition state, determined
by the MZI's first beam-splitter (BS), is uneven
\begin{equation}
\label{eq:pre}
| \Psi \rangle = \frac{4}{5} | R \rangle - \frac{3}{5} | L \rangle.
\end{equation}
Upon exiting the MZI, the atom is post-selected for a special case
with respect to this pre-selection. For example, in contrast to the
first $16/25$--$9/25$ atomic BS within the MZI, let the second BS be
$1/2$--$1/2$. Then post-select the atom for detection events by the
left-hand detector $D$ rather than the normally expected $C$.

Next compute the atom's backwards evolution. It is unitary, hence the
post-selection carries information about the particle's preceding
evolution just as the pre-selection (preparation) gives the evolution
that follows. The retrodiction about the particle's past state within
the MZI is therefore
\begin{equation}
\label{eq:post}
| \Phi \rangle = \frac{1}{\sqrt{2}} | R \rangle + \frac{1}{\sqrt{2}} | L \rangle.
\end{equation}

According to the TSVF, the pre- and post-selected states, though apparently incompatible, are equally valid. Moreover, when
\eqref{eq:pre} and \eqref{eq:post} are inserted into \eqref{eq:wv},
together they give rise to anomalous weak values that have prevailed
during the intermediate time interval within the MZI,
\begin{gather}
\notag
\langle \hat{\Pi}_R \rangle_w = \frac{\left( \langle R | + \langle L | \right)
\hat{\Pi}_R \left( 4 | R \rangle - 3 | L \rangle \right) / \sqrt{50}}
{\left( \langle R | + \langle L | \right) \left( 4 | R \rangle - 3 | L
    \rangle \right) / \sqrt{50}} = 4 \\
\langle \hat{\Pi}_L \rangle_w = \frac{\left( \langle R | + \langle L | \right)
\hat{\Pi}_L \left( 4 | R \rangle - 3 | L \rangle \right) / \sqrt{50}}
{\left( \langle R | + \langle L | \right) \left( 4 | R \rangle - 3 | L
    \rangle \right) / \sqrt{50}} = -3,
\end{gather}
where $\hat{\Pi}_{R / L}$ is a projector on the MZI's right/left arm. Effectively, this is a
description of ``four atoms'' on one arm and ``minus three atoms'' on the
other. Following are two novel measurement techniques that reveal
these weak values.

\subsection{Stimulated Emission indicating the Atom's Odd Weak
  Values}

For validating the above extraordinary weak values, the atom's excited state comes to our aid. Between preparation and post-selection, let the atom be hit by two laser beams (having zero mutual
overlap) directed towards its two possible locations within the MZI
(Fig. 1). Within the Gedankenexperiment's scope, let the interaction
between the photons and the atom be such that it leads to stimulated
emission of radiation with probability approaching 1. Atoms that have
nevertheless remained excited are selected out too. We are therefore
assured that among the photons eventually detected by detectors $L$ and
$R$ there is one additional photon emitted by the excited atom via
stimulated emission.

The laser beams are described by the following coherent state, represented via the real quadrature $q$ (in optical phase space):
\begin{equation}
\psi(q) = (2 \pi)^{-1/2} \exp\left[ - \frac{(q -
    q_0)^2}{4}\right],
\end{equation}
where $q_0 \gg 1$, which is typically the case in the lab.

The interaction with the excited atom is assumed to change this state into:
\begin{equation}
\tilde{\psi}(q) = (2 \pi)^{-1/2} \exp\left[ - \frac{(q -
    q_0')^2}{4}\right],
\end{equation}
where $q_0'=q_0+1/2q_0$ accounts for the additional photon (in suitable units the change in the number of photons is $q_0^2-q_0'^2$ which equals to 1 up to a negligible factor of $1/4q^2_0$).

%(as the unwanted term from the new mean $q_0'^2$ gives rise to $\exp[-1/16q_0^2]\approx 1$).

%For a standard pre-selected only
%ensemble, had $\sigma$ been close to 1, we would have found that one
%photon was emitted from either the left or right location, providing
%which-path information

In the sub-ensemble of atoms under
successful post-selection, the state of emitted photons for the
right arm is \cite{Parks14}
\begin{equation}
\label{eq:em1}
\psi_{\text{em}}^R(q) = \langle \hat{\Pi}_R \rangle_w (2 \pi)^{-1/2} \exp\left[ - \frac{(q -
    q_0')^2}{4}\right] +
\left( 1 - \langle \hat{\Pi}_R
  \rangle_w \right) (2 \pi)^{-1/2} \exp\left[ - \frac{(q -
    q_0)^2}{4}\right],
\end{equation}
where $\hat{\Pi}_R$ is a projector on the MZI's right arm. We have used the
weak value because the state of the atoms is pre- and
post-selected. It is important to note that when $q_0 \gg 1$, the stimulated emission (namely the 1 photon added by the atom
to the initial average number of photons emitted by the source) cannot
provide which-path information. Therefore, although this interaction
between light and matter is ``strong'', the atoms' spatial superposition
barely changes due to the uncertainty in the number of photons. It is this
aspect which makes the proposed technique resemble weak measurement,
yet with the advantage that the atom's energy was strongly measured
and thus allowed observation of an anomalous stimulated emission effect.

Repeat this procedure sufficiently many times to make a reliable statistical estimate of the
number of photons emitted from the atom's two possible locations on
the MZI's two paths.

\begin{figure*}[htb]
\centering

\includegraphics[width=0.6\textwidth]{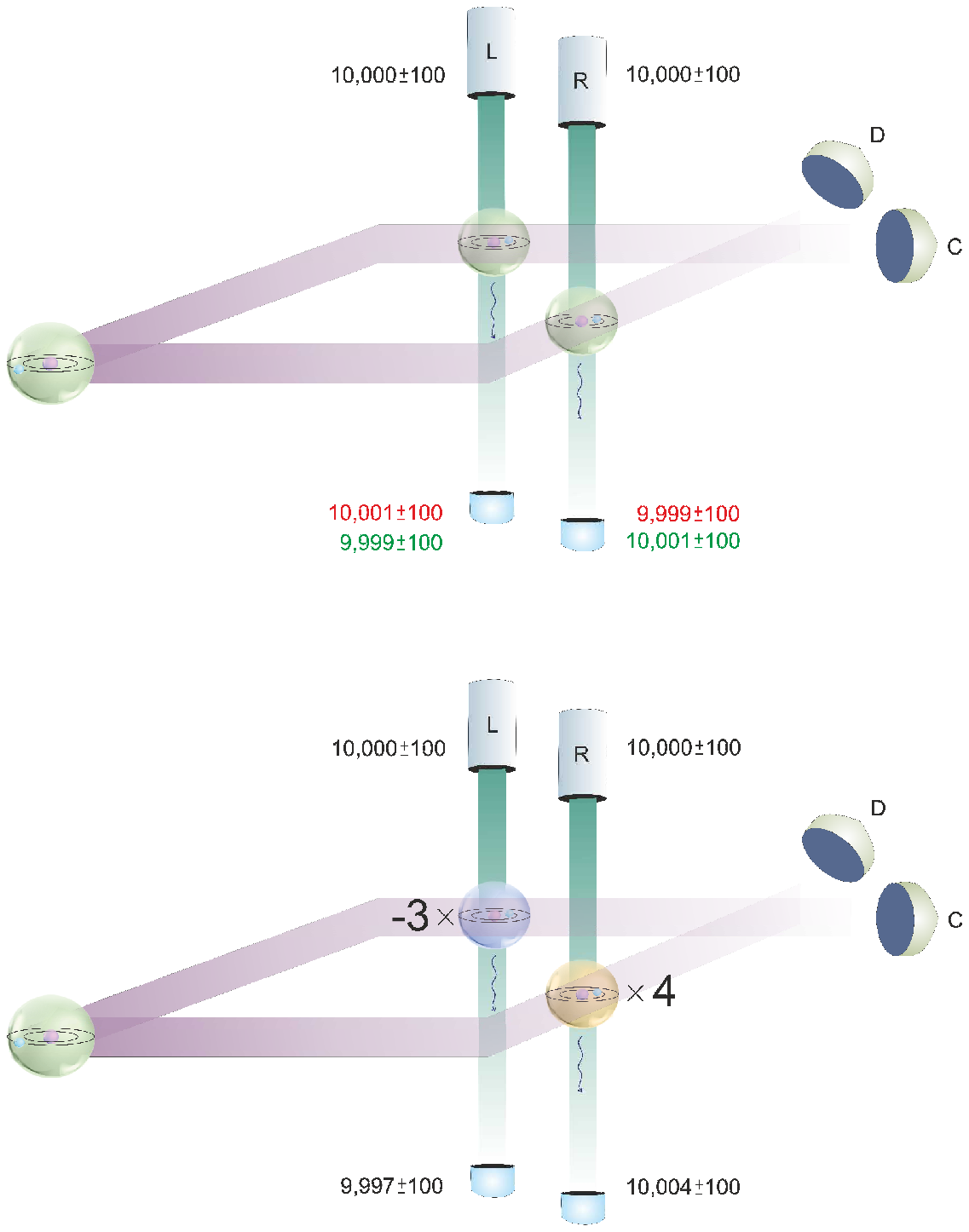}

\caption{An atom (whose state is pre- and post-selected) traverses an MZI and interacts
  with two photonic beams having a large uncertainty in their photon
  number. The excited atom undergoes stimulated emission, becoming
  ground. The number of photons detected at either $L$ and $R$ does not
  give strong ``which-path'' information regarding the atom, hence the atomic
  interference is hardly impaired. Some average numbers have been inserted for
  illustrating the effect. Upper panel: Ordinary post-selection (both
  BSs with 50\% transmission coefficient). The extra photon is
  detected by either $L$ and $R$, making their outcomes correlated yet much below the level of noise. Lower panel: Special pre- and
  post-selection (BS1 with 64\% and BS2 with 50\% transmission
  coefficients). This time the weak values of the particle numbers on
  each arm are -3 and 4. The uncertainty in the number of photons is still greater, but these weak values can be statistically inferred from the average number of hits in each detector using a large enough ensemble.}

\label{fig:a1}
\end{figure*}

Substituting $\langle \hat{\Pi}_R \rangle_w = 4$ in \eqref{eq:em1} and using the first order approximation of the exponential function with $q_0 \gg 1$ , results in
\begin{equation}
\psi_{\text{em}}^R(q) \approx (2 \pi)^{-1/2}\left ( 1-\frac{q-q_0}{q_0} \right)\exp\left[ - \frac{(q -
    q_0)^2}{4}\right] \approx (2 \pi)^{-1/2} \exp\left[ - \frac{(q -
    q_0'')^2}{4}\right],
\end{equation}
where $q_0''=q_0+2/q_0$, i.e. $q_0''^2-q_0^2\approx4$, meaning that in the right-arm position, the atom has reacted to the
laser beam as if it \emph{were 4 atoms} (the new conditional mean has been approximately shifted by 4), emitting 1 photon each. Even
more striking is the similarly derived
\begin{equation}
\psi_{\text{em}}^L(q) \approx (2 \pi)^{-1/2} \exp\left[ - \frac{(q -
    q_0''')^2}{4}\right],
\end{equation}
where $q_0'''=q_0-3/2q_0$, implying that \emph{the effective -3 left-arm-atoms have absorbed one photon each, leading on average to a decrease in the number of photons on this side}.

%This is due to the atom's anomalous
%weak value, which in this case is \emph{negative}. Such an atom, upon
%absorbing a photon, becomes ground, wit no subsequent emission.

As a consistency check, we note that the anomalous numbers of photons $4$ and
$-3$ add up to $1$ as they should. Some numbers have been given in Fig. 1
for illustrating the effect.

To summarize, under a special combination of pre- and post-selections,
the atom traversing the MZI is retrodicted to possess odd
properties. Consider first the familiar superposition $| \Psi \rangle
= (| 1 \rangle + | 2 \rangle) / \sqrt{2}$, that means: upon performing
a projective which-path measurement, there is either $1$ atom in the
right arm or $1$ in the left. Hence, had such an atom absorbed the laser
beam during its passage through the MZI, it would emit one photon from
either side. In our subensemble, however, the atom presents a curious
effective interaction: \emph{There are either $4$ atoms in the right arm or
$-3$  (minus three) atoms in the left, as indicated by the number of
detected photons}.

Generalizing, for the pre-selected state
$$| \Psi \rangle = \left(
  \alpha | R \rangle - \beta | L \rangle \right) / \sqrt{\alpha^2 +
  \beta^2},$$
where $\alpha = \beta + 1 \in \mathbb{R}$, followed by
the post-selected state of \eqref{eq:post}, we would find on average
$\alpha$ extra photons in the right detector and a deficit of $\beta$
photons on the left detector (as long as $q_0 \gg
\alpha$). Similarly, if the superposed atom is ground and we try to excite it, then we expect to find excessive photons on the negative arm.

\subsection{The Atom's Spontaneous Emission of Radiation Appearing
  to Originate from a Phantom Position}

Consider again our excited atom's pre- and post-selected state in
Eqs. 2,3 (Fig. 2). The above predicted appearance of an effective negative
property can be further studied within a complementary scenario, now with
the aid of an interference effect that reveals another unexpected
phenomenon. This time, make the time spent by the atom within the
interferometer much larger than its half-life time, such that between
the pre- and post-selection it is most likely to undergo spontaneous rather than
stimulated emission either at $x=-d$ or $x=d$ (which correspond to the left/right arms of the interferometer, respectively). Make the emitted photon's wavelength $\lambda$ much
larger than the distance between the interferometer arms
$2d$. Consequently, again, radiation cannot reveal which path the atom
took.

\begin{figure*}[htb]
\centering

\includegraphics[width=0.99\textwidth]{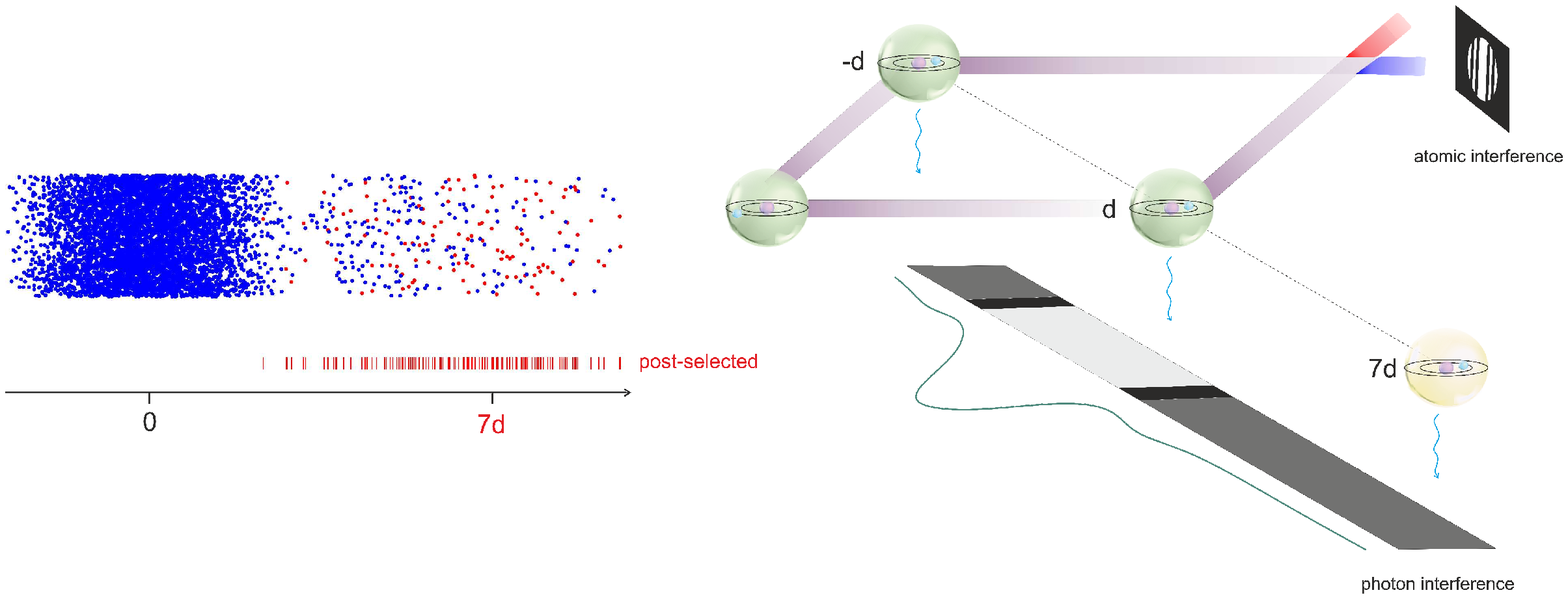}

\caption{An excited atom traversing an atomic MZI, exhibiting
  interference by ending at the blue detector. The atom is delayed within the MZI to allow it to emit a
  photon later absorbed by a photographic plate. The photon's long
  wavelength (in comparison to the size of the MZI) does not allow the presence of ``which path'' information regarding the atom. The experiment
  is repeated sufficiently many times for statistical averaging. The two arms of the MZI are so close to each other compared to the photon's wavelength such that the overall effect is like that of a wide single slit.
Consider the same excited atom under the above special pre- and post-selection,
namely, with 64\% and 50\% transmission coefficients at the first and second beam
splitters followed by detection at the red detector. This time, the
photons detected on the screen, upon statistical averaging, correspond
to a spontaneous emission from an atom effectively located far outside
the interferometer. This, however, is an interference phenomenon of
anomalous weak values.}
\label{fig:a2}
\end{figure*}

Photons emitted from the right/left-arm-atom are described by the following spatial distributions:
$g^{R / L}(x) = \phi(x \pm d)$, having the same functional form, yet
centered around $\pm d$, respectively, with a typical waist of $\lambda$ (for simplicity, $\phi(x)$ can be thought of as being a Gaussian with a standard deviation of order $\lambda$). The photon's long wavelength
allows interference between these two distributions without disclosing
which path information. Let us take advantage of this effect by
detecting these photons with the aid of a photographic plate, just
as in the double-slit experiment. We therefore have two interference
effects, one manifested by the atom itself exiting from the MZI upon
post-selection and the other by the photons it has emitted earlier.

To better comprehend the predicted effect, let us drop for a while the
special post-selection and consider an excited atom traversing an ordinary
MZI while emitting a photon. Here, post-selection is trivial, namely,
exactly reuniting the two atomic wave-function halves split by the
first BS (e.g., both beam splitters being 50\% transparent) and
detecting them in the constructive interference arm C. We expect a single-slit-like pattern, indicating in fact two slits very close to one another (Fig. 2):
\begin{equation}
\phi_{\text{Tot}}(x) = \frac{1}{\sqrt{2}} \left[ \phi(x + d) + \phi(x -
  d) \right].
\end{equation}

Now bring back the above post-selection
\eqref{eq:post}. Statistically averaging again, the total distribution
of photons for the pre- and post-selected cases would be
\begin{multline}
g(x) = \langle \hat{\Pi}_L \rangle_w \phi(x + d) + \langle \hat{\Pi}_R \rangle_w
\phi(x - d)
= -3 \phi(x + d) + 4 \phi(x - d) = [-3 \exp(2i\hat{p}d / \hbar) + 4] \phi(x -
d) \\ \approx (1 - 6 i\hat{p}d / \hbar) \phi(x - d) \approx \exp(-6 i\hat{p}d / \hbar)
\phi(x - d)
\approx \phi(x - 7d),
\end{multline}
where $\hat{p}$ is the photon's momentum and $\phi(x \pm d)$
is the wave emitted from $L$/$R$, respectively. This means that these particular photons exhibit an
interference pattern suggesting that each was emitted by an atom residing in
the ``phantom position'' located at $R' = 7d$ , i.e. far outside the
MZI (see Fig. 2). This is another manifestation of anomalous weak
values, which are now understood to coherently add up via interference
\cite{R3} for yielding this unexpected effect. The latter is akin to
superoscillations \cite{T1,T2} and quantum random walks \cite{T3}. Paradoxically, it is interference which gives rise to this single-slit-like behavior, yet a subtle one based on anomalous weak
values (indeed, had we blocked the photons emerging from either side, this effect would have disappeared).

Generalizing, for the same setup, but with the initial state
$$| \Psi \rangle = \left(
  \alpha | R \rangle - \beta | L \rangle \right) / \sqrt{\alpha^2 +
  \beta^2},$$
where $\alpha = \beta +1 \in \mathbb{R}$, we would observe photons
emitted from the anomalously remote location $(\alpha + \beta)d$ (as
long as $\lambda \gg (\alpha + \beta) d$). Their time of apparent
arrival from $R'$ is therefore similar to that of photons absorbed and re-emitted from locations $L$ and $R$, giving further credence to the effect. We note that there is a tradeoff between the success probability of
the post-selection and the distance of the ``phantom position'' from
the MZI. Superposing the excited atom over $N$ rather than just 2
positions will generally increase the distance at the cost of a lower
success probability. Another case of interest is using imaginary
$\alpha$, $\beta$, which will create a shift in the momentum distribution rather than the
spatial distribution of the emitted photons.

\subsection{Summary}

The above analyses demonstrate the significance of anomalous weak
values in experiments involving interactions between light and
matter. In our case, the anomalous value apparently indicates an atom's negative presence in a certain location, an unusual prediction vindicated with the aid of two thought experiments that reveal two different consequences stemming from it. Importantly, although the light-matter interactions are ``strong'',
the proposed validation techniques, subtly employing quantum
uncertainty, are sensitive to the corresponding weak values.
Quantum superposition, unique in itself, may therefore be understood
as encapsulating some new phenomena emerging upon
post-selection. These in turn lead to the novel concept which merits
further study, namely counter-particles with negative weak values
accompanying particles with positive weak values, together obeying the
familiar conservation laws. \\[4ex]
%
%
%{\bf Data accessibility.} This paper has no data.\\[1ex]
%{\bf Competing interests.} The authors declare no competing
%interests.\\[1ex]
%{\bf Authors' contributions.} Y.A. conceived the main ideas which were later developed by all
%  authors. Y.A. and E.C. performed the mathematical analysis. A.C. performed
%  the simulation and created the figures. A.C.E. and E.C. were the main
%  writers, but all authors contributed to the writing. All authors
%  gave final approval for publication.\\[1ex]
{\bf Acknowledgements.}  We wish to thank Michael Berry and John
Howell for helpful comments.\\[1ex]
{\bf Funding statement.}
Y.A. acknowledges support from the Israel Science Foundation Grant
No. 1311/14, ICORE Excellence Center ``Circle of Light'' and DIP, the
German-Israeli Project cooperation. E.C. was supported by ERC AdG
NLST and by the Canada Research Chairs (CRC) Program. A.C. acknowledges support from Israel Science Foundation Grant
No. 1723/16.\\[1ex]

\end{document}